\documentclass[prl,twocolumn,superscriptaddress]{revtex4}
\usepackage[dvips]{graphicx}

\begin{document}

\draft

\title{Attosecond streaking experiments on atoms: quantum theory versus
simple model}
\author{A. K. Kazansky}
\address{ Fakult\"at f\"ur Physik, Universit\"at Bielefeld,
D-33615 Bielefeld, Germany}
\address{ Fock Institute of Physics, State University of
Sankt Petersburg, Sankt Petersburg 198504, Russia }
\author{N. M. Kabachnik}
\address{ Fakult\"at f\"ur Physik, Universit\"at Bielefeld,
D-33615 Bielefeld, Germany}
\address{ Institute of Nuclear Physics, Moscow State University,
Moscow 119992, Russia }

\date{\today}

%
%

\begin{abstract} A new theoretical approach to
the description of the attosecond streaking measurements of atomic
photoionization is presented. It is a fully quantum mechanical
description based on numerical solving of the time-dependent
Schr\"odinger equation which includes the atomic field as well as
the fields of the XUV and IR pulses. Also a  simple semiempirical
description based on sudden approximation is suggested which agrees
very well with the exact solution.
\end{abstract}

\maketitle


One of the problems in the physics of ultrashort atomic phenomena is
the characterization of the attosecond extreme ultraviolet (XUV)
pulses which are used in pump-probe experiments studying the time
evolution of atomic processes. The characterization includes
measurements of the pulse duration, intensity, basic frequency,
spectral distribution and possible chirp. In a typical experiment
\cite{Hentschel01,Drescher02,Kienberger04,Goulielmakis04}, the XUV
pulses with a duration of a few hundred attoseconds are produced by
filtering out the high energy part of the high harmonic spectrum
which results from a nonlinear interaction of a few-cycle intense
infrared (IR) laser pulse with a gas target. The laser pulse
duration is typically 5-7 fs, its carrier wave length is $\sim 800$
nm (the photon energy $E_L \sim 1.6$ eV), thus the period of the
laser pulse is larger than the XUV pulse duration. The intensity of
the laser  is typically $3\times 10^{13} - 10^{14}$ W/cm$^2$. At
such intensity the electric field of the laser, ${\cal E}_L \sim
(0.1-0.3)\times 10^9$ V/cm, is much smaller than the atomic field (1
a.u.= 5.14 $\times 10^9$ V/cm) although it is strong enough to
produce high harmonics. The electric field of the XUV pulse is still
smaller, about $ 7 \times 10^7$ V/cm \cite{Goulielmakis04}, so that
the linear response approximation can be readily applied to the
interaction of the XUV pulse with atoms. The typical carrier
frequency of the XUV pulse corresponds to the photon energy $E_X
\sim 90$ eV.

The only method of the attosecond pulse characterization, realized
up to now, has been the so-called attosecond streak camera
\cite{Hentschel01,Drescher02,Kienberger04,Goulielmakis04,Constant97,Itatani02,Kitzler02}.
The attosecond pulses under investigation ionize atoms of a gas
target in the field of a co-propagating IR laser pulse. The bunch of
photoelectrons presenting a replica of the XUV pulse interacts with
the field of the few-cycle IR pulse. The streaking effect is
achieved by measuring the photoelectron yield at different delays
between the XUV and laser pulses. The resulting energy and angular
distributions of photoelectrons, bearing information on the original
attosecond XUV pulse, are measured by methods of electron
spectroscopy. Several theoretical methods are used to extract this
information from the photoelectron spectra. They are based on the
classical equation of motion \cite{Itatani02}, the semiclassical
approach \cite{Itatani02,Kitzler02} or quantum mechanical
calculations within a strong field approximation
\cite{Scrinzi01,Kitzler02}. In this way, it has been proved that
single pulses with a duration of as low as 250 as
\cite{Kienberger04} and even 150 as \cite{Wonisch06} have been
produced. The intensity of the XUV pulse has been also measured
\cite{Goulielmakis04}. In addition, the chirp was demonstrated as
being close to zero \cite{Kienberger04}. In spite of the obvious
success of the streak camera method, there are still many problems
in understanding the detailed dynamics of the laser-dressed
photoionization of atoms by the ultra-short pulses, the accuracy and
limitations of the method. The latter problem has been recently
discussed in detail in Ref. \cite{Yakovlev05}.

In this Letter we present fully quantum mechanical calculations of
the double differential cross section of atomic photoionization by
an ultrashort XUV pulse in the presence of a few-cycle intense IR
pulse. The results provide complete information about the process
which may be used either for comparison with an experiment or as a
test-ground for approximate models. We present also a simple model
which permits us to produce a quick analysis of the experimental
data and to study the sensitivity of the measurements to the
parameters of the attosecond pulses.


The photoelectron spectra and angular distributions are calculated
by solving the time-dependent Schr\"odinger equation numerically.
The atom is described by the independent-electron model, and we
assume that the orbitals of all electrons except the active one are
frozen. This approximation is valid for not very strong laser fields
( $ {\cal E}_L \ll 1$ a.u.). We suppose that both IR and XUV photons
are linearly polarized and their polarization vectors are parallel.
In this case there is an axial symmetry with respect to the axis of
polarizations which we choose as $z$-axis. By expanding the active
electron wave function in spherical harmonics
\begin{equation}
\psi (\vec r ,t)=\sum_{\ell=0}^{L_{\max }}u_{\ell}(r,t)Y_{\ell
0}(\theta ,\varphi )
\end{equation}
and substituting the expansion into the time-dependent Schr\"odinger
equation we obtain the following set of equations for the
coefficient functions $u_{\ell}(r,t)$
\begin{eqnarray}
i\frac{\partial }{\partial t}u_{\ell}(r,t) =
-\frac{1}{2}\frac{\partial ^{2}}{ \partial r^{2}}u_{\ell}(r,t)+
\left[ U_\ell(r)+\frac{\ell(\ell+1)}{2r^{2}}\right] u_{\ell}(r,t)
\nonumber
\\ +r {\cal E}_{L}(t)\sum_{\ell'=0}^{L_{\max }}C(\ell,\ell')u_{\ell'}(r,t) \nonumber
\\ + \frac{1}{2} r C(\ell,\ell_0) \delta_{\ell,\ell_0\pm 1}
\bar {\cal E}_{X}(t)\exp [-i(\omega _{X}+\varepsilon_{0})t]
u_{\ell_0}(r). \label{eq:system}
\end{eqnarray}
Here and in the following atomic units are used unless otherwise
indicated. In Eq. (\ref{eq:system}) $U_\ell (r)$ is a single
electron potential, $C(\ell,\ell')$ is the angular part of the
dipole matrix element, $\varepsilon _{0}$ is the binding energy of
the active electron, ${\cal E}_{L}(t)$ is an electric field of the
laser pulse while $\bar {\cal E}_{X}(t)$ is an envelope of the XUV
pulse. Since the XUV field is weak, the first order perturbation
theory and the rotating wave approximation are applied. Then in the
dipole approximation, only transitions $\ell_0 \rightarrow \ell_0
\pm 1$ are possible. In contrast, the laser IR field is strong, it
mixes the states with different orbital angular momenta. In order to
achieve sufficient accuracy, the partial waves up to $L_{max} \sim
40-60 $ have been included.

The system of equations (\ref{eq:system}) has been solved
numerically using the split-propagation method. The details of the
numerical procedure will be published elsewhere (see also Refs.
\cite{Kazan05,Kazan06}). Calculating the set of the functions
$u_{\ell}(r,t)$ at $t \to \infty$ we have evaluated the partial
amplitudes of photoionization by projecting the functions onto the
corresponding continuum functions $\chi_{-}^{(\ell)}(E;r)$:
\begin{eqnarray}
{\cal A}_{\ell}(E ) = \exp [i \eta_{\ell}(E)]
 \int_0^{\infty} d r u_{\ell}(r,t\to+\infty)\chi_{-}^{(\ell)} (E;r) ,
 \label{eq:amplitude2}
\end{eqnarray}
where $\eta_{\ell}(E)$ is the photoionization phase (for details see
Ref. \cite{Kazan06a}). Knowing the amplitudes,  ${\cal A}_{\ell}(E
)$, one can evaluate the double differential cross section (DDCS)
\begin{equation}
\frac{d^{2}{\sigma}(E,\Omega )}{dE d\Omega }= 2 \pi \omega_X \alpha
K^{-1} \left| \sum_{\ell}{\cal A}_{\ell}(E)Y_{\ell
0}(\vartheta)\right|^{2} \,. \label{eq:tripleyield}
\end{equation}
Here $\alpha$ is the fine-structure constant and $K=\int dt |\bar
{\cal E}_{X}(t)|^2$. By integrating Eq. (\ref{eq:tripleyield}) over
the emission angle one can obtain the photoelectron spectrum.

As an example we consider the 3s subshell photoionization in Ar. The
atom is described within the Hartree-Slater approximation
\cite{HS78} with the same self-consistent potential $U(r)$ for each
$\ell$. The laser pulse is assumed to have the form
\begin{equation}
{\cal E}_{L}(t) = \bar {\cal E}_{L}(t)  \cos{[\omega_L
(t-\tau_L)+\phi_L ]}
 \label{eq:IRpulse}
\end{equation}
with the envelope  given by the expression
\begin{equation}
\bar {\cal E}_{L}(t) = {\cal E}_{L0} \frac{1}{2} \{ \cos [\pi
(t/\tau_L-1)]+1 \} \,,
 \label{eq:IRenvelope}
\end{equation}
where $\tau_L$ gives the full width at half maximum (FWHM) and
$\phi_L$ is the carrier-envelope phase. We set $\phi_L = 0$
(cosine-type pulse).
\begin{figure}
\begin{center}
\includegraphics[height=6cm]{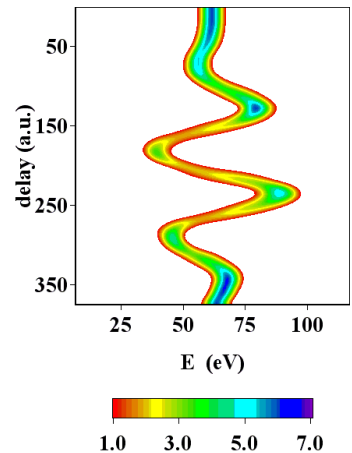}
\end{center}
\caption{(Color online) A 2D plot of the Ar(3s) photoelectron
spectra (electron energy - x-axis) at $0^\circ < \vartheta <
15^\circ $ as a function of the time delay in a.u. (1 a.u. = 24.2
as) between the laser and the XUV pulses (y-axis). The calculations
were performed for the IR pulse of the cosine-type with the duration
$\tau_L = 5.0$ fs, the laser intensity is 10$^{13}$ W/cm$^2$. The
XUV pulse duration is $\tau_X = 250$ as and the carrier frequency is
90 eV. The calculated binding energy of Ar(3s) is 28.6 eV}
\end{figure}
The XUV pulse is delayed with respect to the laser pulse by $t_d$
\begin{equation}
{\cal E}_{X}(t) = \bar {\cal E}_{X}(t-t_d)  \cos{[\omega_X (t-t_d)
]}\,.
 \label{eq:XUVpulse}
\end{equation}
For the envelope $\bar {\cal E}_{X}(t)$ we have chosen the
hyperbolic secant shape. The parameters of the laser and the XUV
pulses are shown in the caption to Fig. 1. This figure shows the
two-dimensional (2D) plot of the calculated spectra as a function of
the delay between the laser and the XUV pulses. The results shown in
Fig. 1 are typical for the streaking experiment
\cite{Drescher02,Kienberger04,Goulielmakis04}.


If the duration of the XUV pulse is much shorter than the period of
the laser field, $\tau_{X}\ll T_L$, then one can consider the whole
process as a two-step one \cite{Itatani02}: first step is
photoionization of the atom by the XUV pulse which is independent of
the laser field, second step is a transport of the emitted electron
by the laser field to the final state. Within the linear response
approximation, the cross section of photoionization by a short pulse
(without the IR field) can be presented as \cite{Kazan06a}
\begin{equation}
\frac{d\sigma _{0}}{d \vec p_{0}}=\frac{a}{\sqrt{2E _{0}}}\left|
D(E_{0})\right| ^{2}\left| f(E_{0}+ | \varepsilon_0 |-\omega_X
)\right| ^{2} Y_{10}^{2} (\vartheta _{0})\,.
\end{equation}
Here $\vec p_{0}$ is the linear momentum of the photoelectron
immediately after photoionization ($\vartheta_0$ is the
corresponding emission polar angle, $E_0$ is the electron energy),
$D(E_{0})$ is the dipole matrix element for transition to the
continuum state with the energy $ E_{0}$, \ $f(\epsilon)$ \ is the
Fourier transform of the XUV pulse at the energy $E_{0}+ |
\epsilon_0 |$, and $a=2\pi\omega_X\alpha K^{-1}$. The factor $Y_{10}
^{2}(\vartheta_{0})$ describes the angular distribution of
photoelectrons for the particular case of the s$\to$p transition
considered here.

With the IR field, the cross section modifies as follows
\begin{equation}
\frac{d\sigma }{\sqrt{2E} dE d\Omega }=\frac{d\sigma }{ d \vec
p}=\frac{d\sigma _{0}}{d \vec p_{0}}\left| \frac{\partial ( \vec
p_{0})}{\partial (\vec p)}\right| ,
\end{equation}
where $\vec p$ is the final momentum of the electron, $\left|
\frac{\partial (\vec p_{0})}{\partial (\vec p)}\right| $ is the
transformation determinant.  Now we are to link the variables
$(\vartheta _{0},E_{0})$ with the experimentally observable
quantities $(\vartheta ,E).$ \ The basic relations for the
components of the momentum $\vec p\equiv \{ p_\parallel, p_\perp \}$
parallel and perpendicular to the field are
\begin{eqnarray}
p_{\parallel } = p_{0\parallel }-\int_{\tau }^{\infty}{\cal E}_L
(t)~dt\equiv p_{0\parallel }- A_L(\tau), \label{eq:parallel} \\
\quad p_{\perp }=p_{0\perp },
\end{eqnarray}
from which it follows $\left| \frac{\partial (\vec p_{0})}{\partial(
\vec p)} \right|=1$ and
\begin{eqnarray}
\sin \vartheta _{0} = \sqrt{E/E_{0}}\sin \vartheta
\,.\label{eq:sinus}
\end{eqnarray}
Here $\tau$ is the time of the photoelectron emission  and $A_L
(\tau) $ is a parallel component of the vector potential of the
laser field in the Coulomb gauge. From Eqs. (\ref{eq:parallel}) and
(\ref{eq:sinus}) it follows that
\begin{equation}
E_{0}= E +\sqrt{2E} A_L (\tau) \cos \vartheta +A_L^{2}(\tau) /2.
\label{eq:E0E}
\end{equation}
The final approximate expression for the DDCS reads:
\begin{equation}
\frac{d\sigma }{dE d\Omega }= a \,\sqrt{\frac{E }{E_{0}}}\left|
D(E_{0})\right| ^{2}\left| f(E_{0}+| \varepsilon_0|-\omega_X)\right|
^{2} Y_{10}^{2}(\vartheta_{0}) \,, \label{eq:final}
\end{equation}
where $E_0, \vartheta_0$ and $E, \vartheta$ are connected by Eqs.
(\ref{eq:E0E}, \ref{eq:sinus}).

In Fig. 2(a) we compare the DDCS calculated by the numerical
solution of the Schr\"odinger equation and by using Eq.
(\ref{eq:final}) in the case when the delay time is 7 fs.
\begin{figure}
\begin{center}
\includegraphics[height=6cm]{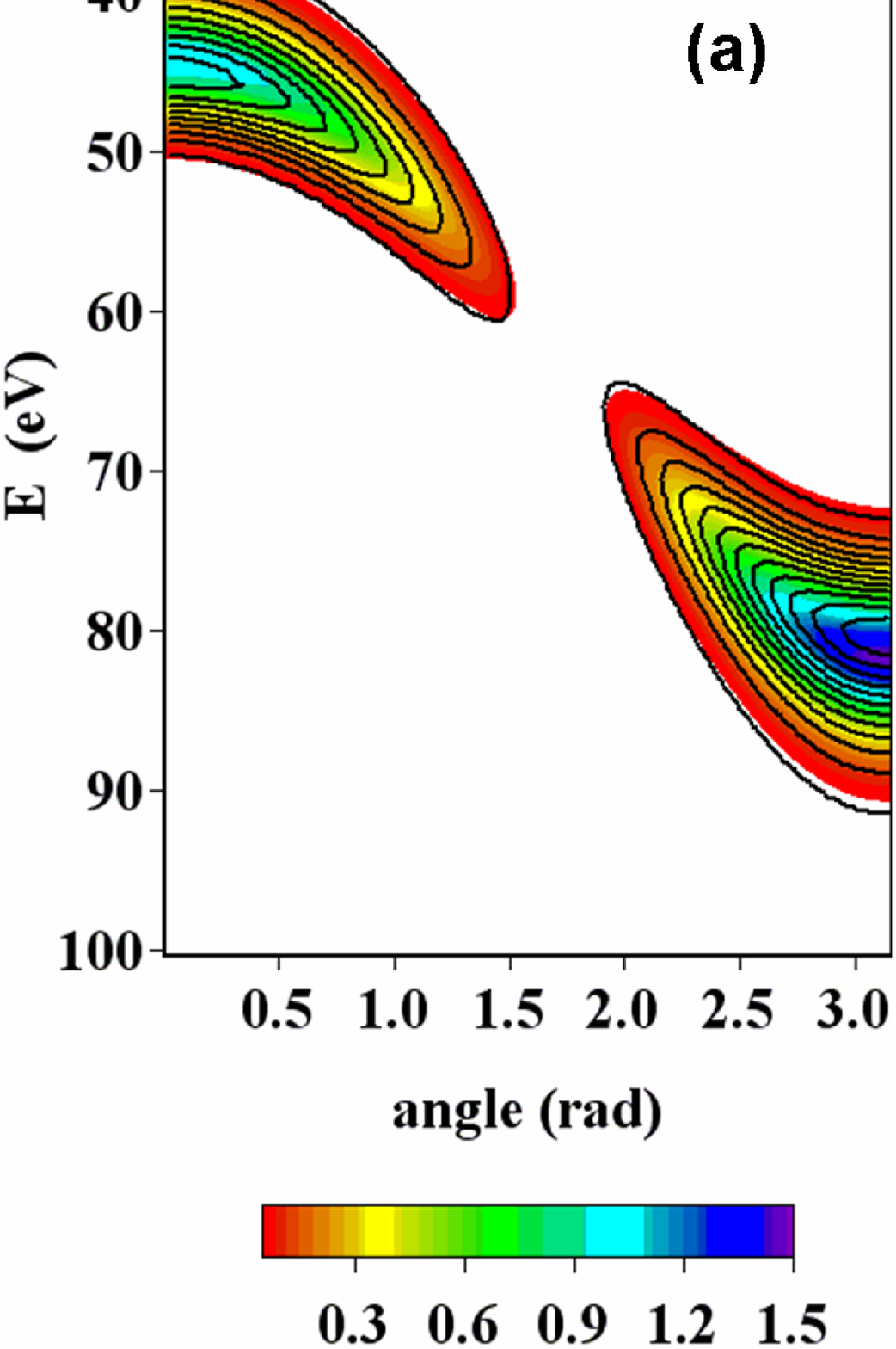}
\includegraphics[height=6cm]{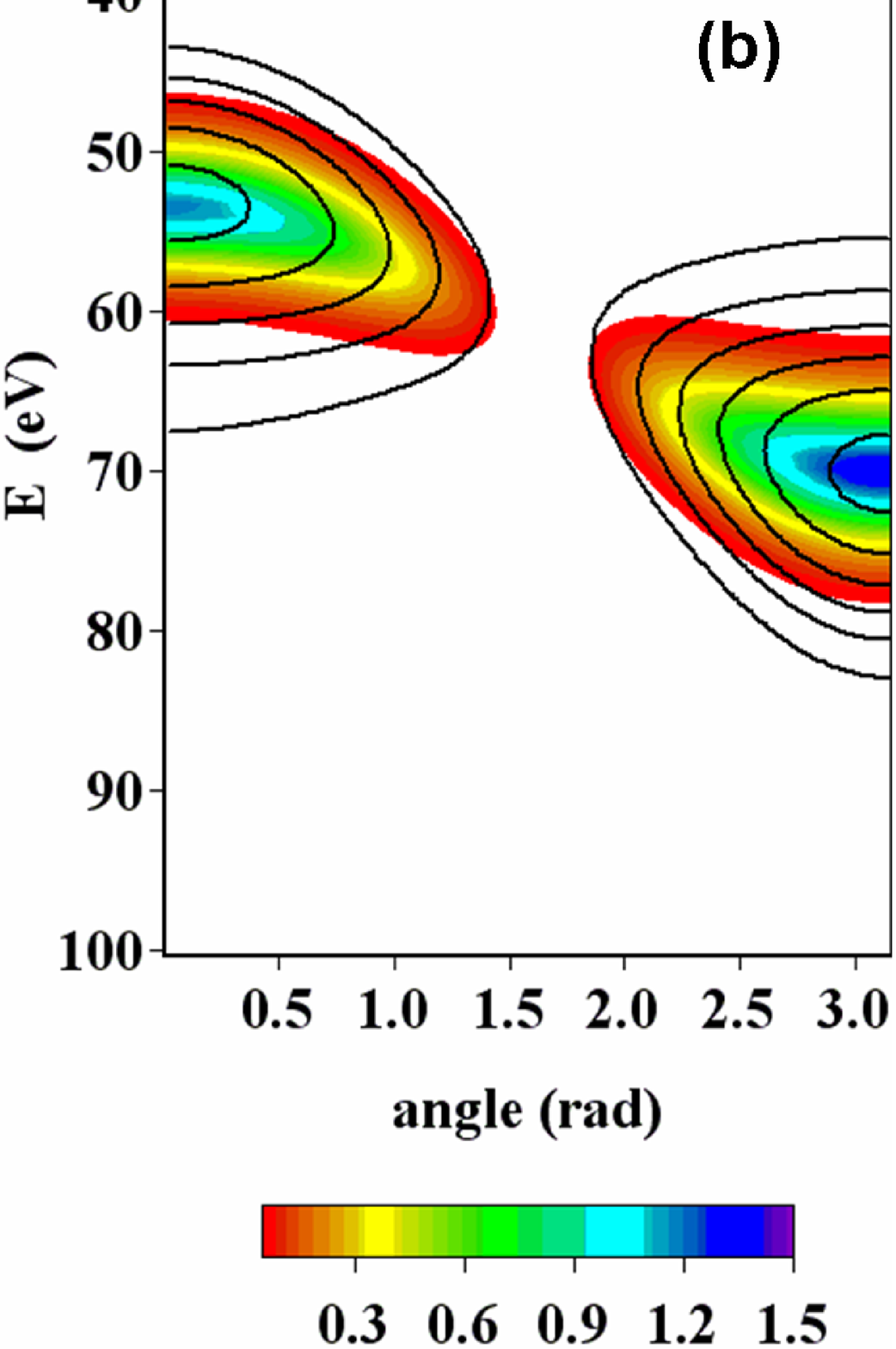}
\end{center}
\caption{(Color online) 2D plots of the calculated DDCS for Ar(3s)
photoionization for different time delays between the laser and the
XUV pulses: (a) $t_d= 290$ a.u. (7 fs), (b) $t_d= 270$ a.u. (6.5
fs). The parameters of the pulses are the same as in Fig. 1. Contour
plots show the results obtained by solving the Schr\"odinger
equation; color (grey) scale plots show the results of model
calculations using Eq. (\ref{eq:final}).  The minimal and maximal
values are the same for both plots. }
\end{figure}
The agreement between the two calculations is almost perfect. Note
that at the chosen delay ${\cal E}_L(t_d)$ is almost zero. Similar
agreement is obtained for all other delays when the laser field is
close to zero. In contrast, for the case of $t_d = 6.5$ fs shown in
Fig. 2(b), the distribution given by the exact computation is
substantially broader than that given by the approximate formula.
Similar discrepancy is obtained for other delays when the value of
the laser field is close to its extrema.

From the angular distributions calculated for each time delay we
obtain the photoelectron energy spectra at a given angle by
integrating over some acceptance angle ($15^\circ$) of the detector.
These spectra are fitted with the Gaussian function
$A\,\exp[-(E-\bar E)^2/\Gamma^2]$ in order to determine their
widths. Variation of the spectral width with the time delay is
commonly used in order to determine the parameters of the XUV pulses
\cite{Hentschel01,Kienberger04}. In Fig. 3 we show the widths
$\Gamma$ of the calculated spectra as a function of the delay time
for two angles $\vartheta = 0^\circ$ and 180$^\circ$. Comparing the
results of the exact calculations (symbols) with the calculations
using the approximate formula Eq. (\ref{eq:final}) (dotted curves)
we have noticed that the difference is always proportional to the
value of the IR  laser electric field at the moment of the
photoelectron emission. As is known \cite{Itatani02}, the broadening
of the electron spectra (for electrons emitted in the same
direction) is determined by two factors. First, the spectrum becomes
broader since the emitted electrons have different energies due to
the large spectral width of the XUV pulse. This type of broadening
is taken into account by Eqs. (\ref{eq:E0E}) and (\ref{eq:final}).
Second, the electrons are emitted at different times within the XUV
pulse duration and, therefore, at slightly different phases of the
IR field that leads to an additional spread of the electron
energies. Since $\tau_X \ll T_L$,  this spread depends linearly on
the field at the moment of electron emission, and in the first
approximation it is proportional to the duration of the XUV pulse.
Thus we suggest a semiempirical expression for the width:
\begin{equation}
\Gamma = \Gamma_m + \sqrt{2E_0}{\cal E}_L(t_d)\tau_{1/2}/2 \,,
\label{eq:empirical}
\end{equation}
where $\Gamma_m$ is a model width, calculated from the spectra
obtained using Eq. (\ref{eq:final}) and $\tau_{1/2}$ is a FWHM of
the XUV field
envelope.
Using this expression we obtained the values, shown
by solid lines in Fig. 3, which are in excellent agreement with the
results of the exact calculations shown by open symbols, circles
(forward emission) and squares (backward emission).

\begin{figure}
\begin{center}
\includegraphics[height=4cm]{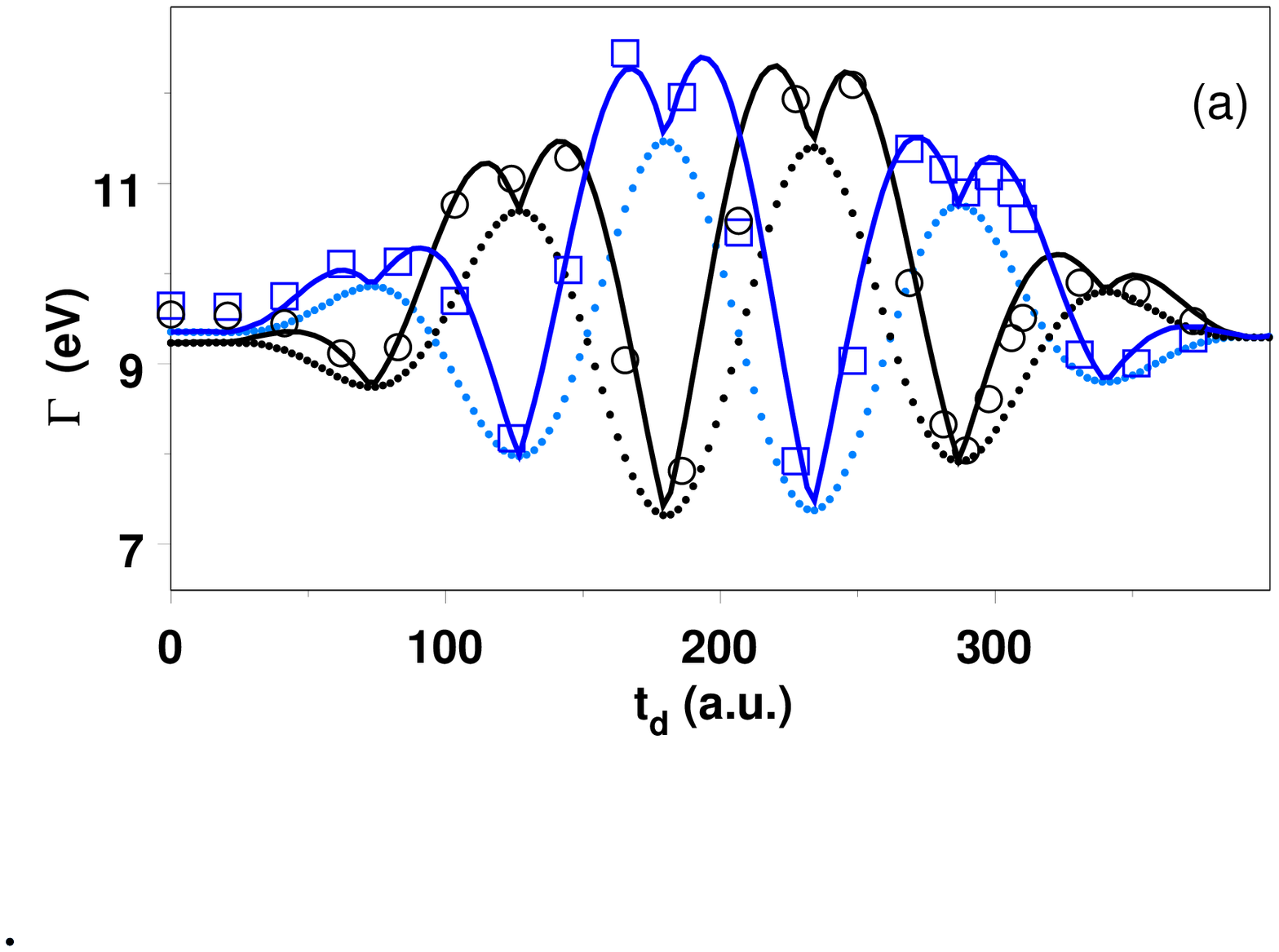}
\includegraphics[height=4cm]{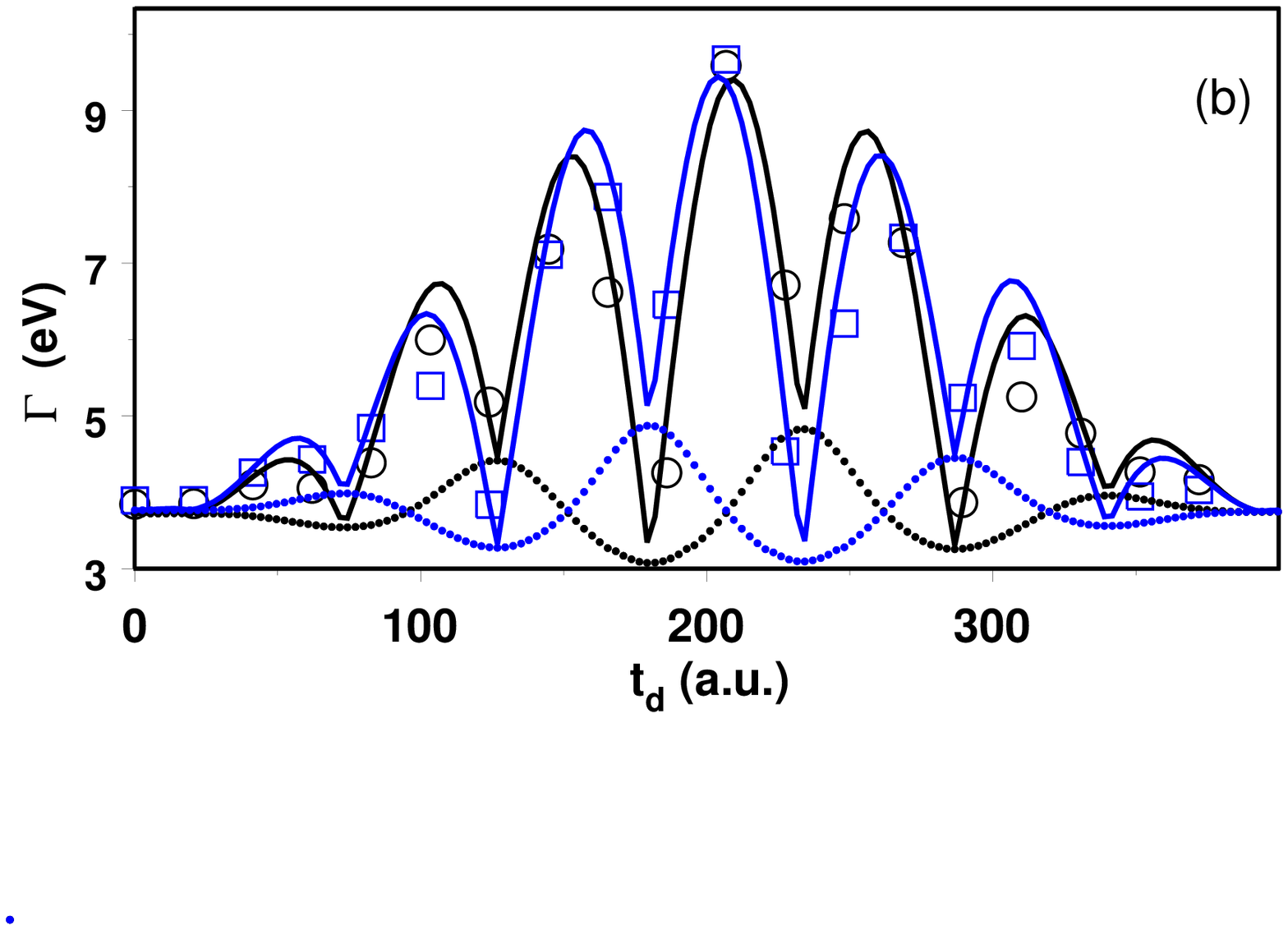}
\end{center}
\caption{(Color online) The width of the photoelectron spectrum as a
function of the delay time for two XUV pulse durations: (a)
$\tau_X=100$ as, (b) $\tau_X=250$ as. Black symbols and curves
correspond to forward emission $\vartheta=0^\circ$, blue (gray)
symbols and curves correspond to backward emission
$\vartheta=180^\circ$. Exact calculations: circles
($\vartheta=0^\circ$) and squares ($\vartheta=180^\circ$). Dotted
curves show $\Gamma_m$ obtained by model calculations. Solid curves
show the results obtained by Eq. (\ref{eq:empirical}). All
parameters in the calculations are the same as in Fig. 1.}
\end{figure}

Besides very good agreement between the calculations, an interesting
feature can be seen in Fig. 3, namely a double-maximum structure in
the resulting model curve for the pulse duration of 100 as. This is
not an artefact since the exact calculations with small steps in
time delay around $t_d = 300$ a.u. perfectly agree with the model
calculations. The structure is due to the fact that at the
zero-crossing points where ${\cal E}_L(t_d)=0$, the width should be
equal to that obtained from Eq. (\ref{eq:final}), and therefore a
minimum is formed. The depth of the minimum is determined by the
contribution of the pulse spectral width (dotted curve). For longer
pulses (250 as) this contribution is smaller and the minimum is
deeper. An analysis of the results obtained with model expression
Eq. (\ref{eq:final}) shows that for very short XUV pulses ( $\le$
100 as) the model describes the width quite well, as well as
describing the whole structure of the DDCSs.  Thus, for such short
pulses one can keep track only on the spectrum of the XUV pulse but
not on its duration and the chirp. With an increase of the XUV pulse
duration, the first term in Eq. (\ref{eq:empirical}) is decreasing
while the second one is increasing and becomes dominant. The second
term gives information on the FWHM of the XUV envelope. The latter
quantity can also depend on the chirp, although this dependence is
not strong. Thus, the chirp can be obtained only for the pulses for
which the second term in Eq. (\ref{eq:empirical}) is sufficiently
large. The value of this term is proportional to the strength of the
IR field and to the duration of the XUV pulse. Thus stronger IR
fields and longer XUV pulse durations are favorable for chirp
registration.


In conclusion, we have developed a fully quantum mechanical theory
of the streaking measurements in the attosecond region based on
numerical solving of the time-dependent Schr\"odinger equation which
includes the electron interaction with the atomic core as well as
with the XUV and the laser fields. We suggest also a simple model
expression for the width of the photoelectron spectra which agrees
very well with the exact calculations. Using this expression it is
easy to analyze the sensitivity of the width to the parameters of
the XUV pulse.

The financial support from the Russian Foundation for Fundamental
Research via Grants 05-02-16216 and 06-02-16289 is gratefully
acknowledged.

\newpage

\end{document}